\documentclass[preprint]{elsarticle}
\usepackage{cmap}
\usepackage[T2A]{fontenc}
\usepackage[utf8]{inputenc}
\usepackage[english]{babel}
\usepackage{siunitx}
\usepackage{graphicx}
\usepackage{amsmath}
\usepackage{amssymb}
\usepackage{url}
\usepackage{tikz}
\usetikzlibrary{arrows}
\usetikzlibrary{trees}

\newcommand{\Xdef}[1]{\emph{#1}}

\begin{document}
\begin{frontmatter}

\title{Astronomical observation tasks short-term scheduling using PDDS algorithm}

\author[myaddress]{Matwey V. Kornilov}
\ead{matwey@sai.msu.ru}
\address[myaddress]{Sternberg Astronomical Institute, Lomonosov Moscow State University, Universitetsky~pr.,~13, Moscow~119991, Russia}

\begin{abstract}
A concept of the ground-based optical astronomical observation efficiency is considered in this paper.
We believe that a telescope efficiency can be increased by properly allocating observation tasks with respect to the current environment state and probability to obtain the data with required properties under the current conditions.
An online observations scheduling is assumed to be an essential part for raising the efficiency.
The short-term online scheduling is treated as the discrete optimisation problems which are stated using several abstraction levels.
The optimisation problems are solved using the parallel depth-bounded discrepancy search~(PDDS) algorithm~\cite{Moisan2014}.
Some aspects of the algorithm performance are discussed.
The presented algorithm is a core of open-source {\tt chelyabinsk} C++ library which is planned to be used at~$\SI{2.5}{\meter}$~telescope of Sternberg Astronomical Institude of Lomonosov Moscow State University.
\end{abstract}


\end{frontmatter}

\section{Introduction}
Since an efficiency is a philosophic concept, it is impossible to give it an unique and precise definition both in general and in the particular case of astronomical observations.
Even when ground-based optical astronomy is considered, different concepts are used as an efficiency.
In case of dedicated small robotic observatories, open-shutter time is considered as a measure of an efficiency.
A fast cadence is desired when surveys are performed.
More classical definition by Bowen~\cite{Bowen1964} assumes that efficiency is related to the limiting magnitude of a telescope.
In other words, it is assumed that unexplored and challenging targets belong mostly to the faint object area.
In some sense, this assumption is still valid today. 

Further, we accept Bowen point and try to develop this idea.
We consider a set of an atmosphere, an optical system and an equipment as a single physical system used for carrying experiments (astronomical observations in our case).
Modern ground-based astronomical observations are affected by different external factors, for instance, an atmospheric optical turbulence is commonly mentioned as a phenomenon limiting optical angular resolution.
Effect of the optical turbulence doesn't remain the same but constantly changes over the time.
We may consider the physical system evolution as a point in a phase space, where each axis corresponds to a physical quantity affecting astronomical observations.
The physical quantities are divided into different groups.
Those which don't vary significantly over the time: a telescope aperture size, a CCD readout noise, etc.
The quantities which are under our control, for instance, equipment settings or a telescope mount position.
The last part is the quantities which are not under control: an atmospheric optical turbulence power, an atmospheric extinction, a night sky brightness, etc.
In other words, the system evolves stochastically over corresponding axes.

It is assumed that the system is in the particular area of the phase space during classical ground-based astronomical observations of a specific target.
For instance, to carry out separate photometry of a binary star with the separation of $\SI{1.4}{\arcsecond}$, we have the reasons to demand that the optical resolution should be well better than $\SI{0.7}{\arcsecond}$.
For each particular observation task a feasible area has different size and form.
Even more, time resources of almost any modern general-purpose optical telescope are limited.
Different scientific tasks and programs have to compete with each other for available resources.

The astronomical observation scheduling concept is usually divided into a long-term scheduling and a short-term one.
The long-term scheduling considers time ranges of days, weeks, or months.
It may use some statistical information about environment, but the long-term scheduling is not required to be performed online.
The short-term scheduling considers ongoing night and is usually thought as of online procedure using live data about environment~\cite{Gomez2003}.
Only short-term online scheduling is considered further in the paper.

The short-term online scheduling is supposed to raise an efficiency at least by avoiding idles due to unfeasible conditions.
We essentially follow the idea behind Bowen formula that supposes a telescope can be considered to be more efficient than another one if more observation tasks can be carried out within the same time interval~(and time resources are left for more observations).

We assume that for the upcoming night there is a task set generated by a long-term scheduling process~(either automatic or manual).
For any particular time moment of an ongoing night we want to select an ordered subset of tasks to observe right now and in the near future.
It is assumed that the subset is selected in globally effective way.
We don't consider further what happens with the tasks that have not been selected and have not been observed.
However, the most obvious way would be to return the tasks to the long-term scheduler.

\section{Optimisation problems}
\label{opt_prob}
As soon as we talk about automatic scheduling (i.e. a kind of algorithm in generic sense) a concept of efficiency has to be operationalised in specific way.
A variety of astronomical observational tasks (and scientific knowledge) is to be reduced to a single number.
Definitely, it can't be done uniquely and precisely.
Nevertheless, the following quantities are introduced.

Let $T$ be a set of all available observational tasks.
For any observational task $i \in T$ let $p_i(t|\theta)$ be conditional success probability viewed as a function of task observation start time moment~$t$.
Here the current system state (which is the system state history in essence) is denoted by $\theta$ and will be skipped in the further equations for brevity.
An observation task is said to be successfully carried out when the system is in appropriate area of the state space during observation of the task.
We assume that the system trajectory in the state space can be somehow forecasted given that the current state $\theta$ is known.
The state is supposed to be known by means of dedicated monitoring systems~\cite{Colome2010} or by means of online observation processing pipelines~\cite{Delgado2014}.
A relative weight of observational task is called a \Xdef{yield} and is denoted by $y_i(t)$.
The set $T$ is considered to be finite, then without loss of generality, it can be assumed that $0 \le y_i(t) \le 1$.
A set of all non-empty finite sequences consisted of members of $T$ is denoted by $T^{+}$.
Let $S \in T^{+}$ be a non-empty finite task sequence, further we assume that $\forall i \neq j, S_i \neq S_j$.
The number of elements in $S$ is denoted by $|S|$.

Finally, a total yield is defined as the following:
\begin{equation}
\label{stochastic_yield}
{\mathcal Y} = \sum_{i=1}^{|S|} y_{S_i}(t_{S_i}) \xi_{S_i}(t_{S_i}),
\end{equation}
where $\xi_{S_i}(t_{S_i})$ are random binary variables being~$1$ with probability of $p_{S_i}(t_{S_i})$.
All $\xi_i$ are assumed to be independent for the sake of simplicity.
$t_{S_i}$ are introduced in the following recurrent manner:
\begin{equation}
\label{def_t}
t_{S_{i+1}} = t_{S_i} + d_{S_i}(t_{S_i}) + s_{S_i,S_{i+1}}\left(t_{S_i} + d_{S_i}(t_{S_i})\right),
\end{equation}
where $t_{S_1}$ is the initial time moment.
Without loss of generality, one may assume that $t_{S_1} = 0$.
$d_{S_i}(t)$ is a duration of task observation process when started at~$t$,
$s_{S_i,S_{i+1}}(t)$ denotes a setup time required to start task $S_{i+1}$ after task $S_{i}$ has been completed.
The mean of~(\ref{stochastic_yield}) is called a mean total yield:
\begin{equation}
\label{mean_yield}
Y \equiv {\mathrm E}\left[{\mathcal Y}\right] = \sum_{i=1}^{|S|} y_{S_i}(t_{S_i}) p_{S_i}(t_{S_i}),
\end{equation}
Note that the total yield is the weighted number of successfully completed observational tasks in essence. 

By the previous assumptions, the probability of finite task sequence $S$ success is the following:
\begin{equation}
\label{prob}
\Pi = \prod_{i=1}^{|S|} p_{S_i}(t_{S_i}).
\end{equation}

Let us state two following discrete optimisation problems which are considered further as observational scheduling problems.
Then, mean total yield maximisation problem is
\begin{equation}
\label{max_mean_yield}
Y^* = \max_{S \in T^{+}} \left( \sum_{i=1}^{|S|} y_{S_i}(t_{S_i}) p_{S_i}(t_{S_i}) \right).
\end{equation}
Success probability maximisation problem is
\begin{equation}
\label{max_prob}
\Pi^* = \max_{S \in T^{+}} \left(  \prod_{i=1}^{|S|} p_{S_i}(t_{S_i}) \right).
\end{equation}
Constraints for sequence length are provided for both of the problems.
In the first case:
\begin{equation}
\label{max_mean_yield_cst}
t_{S_{|S|}} + d(t_{S_{|S|}}) \le D,
\end{equation}
in the second case:
\begin{equation}
\label{max_prob_cst}
t_{S_{|S|}} \ge D,
\end{equation}
where $D$ has a sense of scheduling horizon or sunrise moment.
The function $S^*(\theta)=\arg\max_{S \in T^{+}} \left( \sum_{i=1}^{|S|} y_{S_i}(t_{S_i}) p_{S_i}(t_{S_i}|\theta) \right)$ (and its analogue for case~(\ref{max_prob})) is also usually called as decision process a-priory policy.

Therefore, we connect a concept of ground-based astronomical observations efficiency with the yield in~(\ref{max_mean_yield}),
or with success probability in~(\ref{max_prob}).
The problems are complementary in some sense.
The number of successes are maximised in~(\ref{max_mean_yield}) and the number of failures are minimised in~(\ref{max_prob}).
This quantities are based on some natural concepts~(i.e. number of performed tasks) and replicate existing models~\cite{Gomez2003} in some sense.

Let us again emphasise that there is a crucial logical gap between philosophical concept and any its specific numerical measure.
Thus, instead giving ultimate formal proof of equivalence between a concept and its measure, we can consider the measure only as a representation for the concept.
It is for end users to decide whether the particular measure is relevant to the concept.
The decision is based on current understanding what the telescope efficiency concept really is under particular circumstances.
Moreover, the understanding will inevitable be changed as gaining practical experience.
Therefore, our approach should be flexible enough to be modified in future with new demands.

Consequently, it is also impossible to determine which approach (mean total yield maximisation problem~(\ref{max_mean_yield}) or success probability maximisation problem~(\ref{max_prob})) is the most right one, because
the comparison is possible only on philosophical or methodological levels, which is behind the scope of this paper.
Indeed, let $S^*_1$ and $S^*_2$ be solutions for~(\ref{max_mean_yield}) and~(\ref{max_prob}) respectively.
Also, let $f$ be a metric such that the higher value $f(S^*_{1,2})$ the more correct and more adequate the problem has been formulated.
Then~(\ref{max_mean_yield}) and~(\ref{max_prob}) are to be considered as approximations to the maximisation problem of $f(S)$ which is actually being solved and $f$ is implicitly considered as another efficiency measure.



\subsection{Forms of $p(t)$, $d(t)$, $s(t)$}

Let us consider possible forms of the functions $p(t)$, $d(t)$, and $s(t)$ from~(\ref{max_mean_yield}) and~(\ref{max_prob}).
Also it will become more clear what we assume as an abstraction called an observational task.
All tasks of $T$ may have different origin, but the functions $p(t)$, $d(t)$, and $s(t)$ form an abstraction level between physical model and the optimisation problem.
Further we consider different kinds (or classes) of observational tasks: a group, a repeat, CCD-based photometry task.

\subsubsection{Group}
\label{sss_group_task}
A task of the group class is in essence an ordered finite task sequence denoted here by $S$.
Let $g$ be a group task, then the functions are expressed in the following way:
\begin{equation}
\label{group_task_yield}
y_g(t) = \sum_{i=1}^{|S|} y_{S_i}(t + t_{S_i}),
\end{equation}
\begin{equation}
\label{group_task_prob}
p_g(t) = \prod_{i=1}^{|S|} p_{S_i}(t + t_{S_i}),
\end{equation}
\begin{equation}
\label{group_task_duration}
d_g(t) = t_{S_{|S|}} + d_{S_{|S|}}(t + t_{S_{|S|}}),
\end{equation}
\begin{equation}
\label{group_task_setup1}
s_{g,j}(t) = s_{S_{|S|},j}(t),
\end{equation}
\begin{equation}
\label{group_task_setup2}
s_{j,g}(t) = s_{j,S_{1}}(t).
\end{equation}
Note that we again don't specify the origin of task $S_i$.
There may be a group of groups.
One of possible use-cases could be multiband CCD-photometry task of an object,
when the group would be a sequence of CCD~exposures of the same target with different filters.
A group is atomic or non-preemptive.

\subsubsection{Repeat}
\label{sss_repeat_task}
A task of the repeat class is defined as a finite task sequence consisted of $N$ subsequent copies of the task $\tau \in T$.
The parameter $N$ is specified either directly or by using constraints.
The execution duration of a repeat task is expressed recursively in the following way:
\begin{equation}
\begin{split}
d(t,N) &= d(t,N-1) + s_{\tau,\tau}\left(t+d(t,N-1)\right) \\ 
       &+ d_{\tau}\left(d(t,N-1)+s_{\tau,\tau}(t+d(t,N-1))\right),
\end{split}
\end{equation}
where $d(t,1) \equiv d_{\tau}(t)$.
The probability that there are at least $K$ successes given $N$ repeats:
\begin{equation}
p(t,K,N) = p_{\tau}(t_N) p(t_{N-1},K-1,N-1) + \left(1-p_{\tau}(t_N)\right) p(t_{N-1}, K, N-1),
\end{equation}
where $p(t,0,N) \equiv 1$, $p(t,K,0) \equiv 0 (K \neq 0)$, $t_N \equiv t + d(t,N)$.
Corresponding mean yield:
\begin{equation}
\begin{split}
y(t,K,N) &= p_{\tau}(t_N) y_{\tau}(t_N) p(t_{N-1},K-1,N-1) \\
	&+ p_{\tau}(t_N) y(t_{N-1},K-1,N-1) + \left(1-p_{\tau}(t_N)\right) y(t_{N-1}, K, N-1),
\end{split}
\end{equation}
where $y(t,K,0) \equiv 0$, $y(t,0,N) = p_{\tau}(t) y_{\tau}(t) + y(t,0,N-1)$.

If $N$ is not specified directly; then the following constraints are considered in order to specify $N$ indirectly.
Firstly, the probability that there are at least $K$ successes given $N$ repeats is greater than $p_0$:
\begin{equation}
\label{repeat_task_cst_prob}
p(t,K,N) \ge p_0,
\end{equation}
This constraint is trivial when $p_0=0$.
Substituting $N-K$ for $K$ in~(\ref{repeat_task_cst_prob}) we obtain the probability that there are at most $K$ failures given $N$ repeats.

Secondly, the repeat duration is greater (or less) than specific value $D$:
\begin{equation}
\label{repeat_task_cst_duration}
d(t,N) \ge D, \quad(\mbox{or}\, d(t,N) \le D)
\end{equation}
The greater-than constraint is always satisfied when $D=0$.
Changing $d(t,N)$ to $t + d(t,N)$ in~(\ref{repeat_task_cst_duration}) an absolute time constraint is introduced.
The constraint requires that the task ends later (or earlier) than specific time moment.

Now we can declare two different optimisation problems to find the repeat count $N$.
When $N$ is found as a solution for the problem
\begin{equation}
N = \max_{N \in {\mathbb N}} N,
\end{equation}
with optional constraints of form~(\ref{repeat_task_cst_prob}) and~(\ref{repeat_task_cst_duration}), we call it a \Xdef{greedy repeat}.
Similarly, we call it a \Xdef{lazy repeat}, when $N$ is found from the problem:
\begin{equation}
N = \min_{N \in {\mathbb N}} N,
\end{equation}
with the same constraints.

Trivially, the setup time $s(t)$ coincides with the setup time of $\tau$.

The considered abstraction follows from the requirement to monitor specific astronomical targets during specific time window.

As well as a group, a repeat is also atomic or non-preemptive.
Unfortunately, it restricts formulating certain kinds of tasks.
For instance, if we want to monitor a target once per hour but need only a few minutes to perform its observation.
To construct preemptive counterparts of groups and repeats we would need inter-task dependencies.
They are certainly not a must-have feature and not considered in this paper.
However, if described framework were found acceptable in practice then dependencies would be introduced further.

\subsubsection{CCD-based photometry}
\label{ccd_based_photometry}
Let us consider now a task class related to hardware equipment, a CCD-based astronomical photometer.
We assume that the device is carrying out an exposure of the specified sky target with requested characteristics.
A spectral pass band, binning, readout mode may be considered among the user-defined input task parameters.
We consider the following goal requirements: required relative photometric error $\epsilon$, required exposure time $\tau$, central intensity~$\gamma_1$ of the point spread function~(PSF), full width at half maximum~(FWHM)~$\gamma_2$, radius~$\gamma_3(e)$ encircling part of energy~$e$~\cite{Kornilov2015}.
For any quantity from the list an appropriate constraint may be formulated, some (or all) quantities may be left unconstrained.
A set of the constraints and the required device state (including spectral pass band) form CCD-based photometry task abstraction.

Depending on the set of the constraints, functions $d$ and $p^{0}$ are defined as the following.
If the exposure time $\tau$ is specified and the photometric error $\epsilon$ is not; then:
\begin{equation}
\label{ccd_task_duration1}
d(t) \equiv \tau,
\end{equation}
\begin{equation}
\label{ccd_task_prob1}
p^0(t) \equiv 1.0.
\end{equation}
If both the exposure time $\tau$ and the relative photometric error $\epsilon$ are specified; then
\begin{equation}
\label{ccd_task_duration2}
d(t) \equiv \tau,
\end{equation}
\begin{equation}
\label{ccd_task_prob2}
p^0(t) \equiv F_{\epsilon}(\epsilon),
\end{equation}
where  $F_{\epsilon}(\epsilon)$ is a cumulative distribution function~(CDF) of the photometric error forecast given the exposure time $\tau$.
If the exposure time $\tau$ is not specified and the photometric error $\epsilon$ is; then
\begin{equation}
\label{ccd_task_prob3}
p^0(t) \equiv 0.95.
\end{equation}
\begin{equation}
\label{ccd_task_duration3}
d(t) \equiv Q_{\tau}(p^0(t)),
\end{equation}
where $Q_{\tau}(p)$ is a quantile of the required exposure time $\tau$ forecast distribution.
The complete success probability is
\begin{equation}
\label{ccd_task_prob}
p(t) = p^0(t)\left(1-F_{\gamma_1}(\gamma_1)\right)F_{\gamma_2}(\gamma_2)F_{\gamma_3}(\gamma_3),
\end{equation}
where $F_{\gamma_i}(\gamma_i)$ are CDFs of $\gamma_i$ obtained in the paper~\cite{Kornilov2015}.
Note that $F_{\gamma_i}(\gamma_i)$ are derived from a seeing (an angular size of point spread function due to atmospheric optical turbulence) forecast.
The forecast is represented as a multivariate conditional probability density function for the seeing $\beta_i$, where $\beta_i$ are taken at 1-minute intervals.
The probability density function is calculated using on-line seeing monitor measurements and autoregressive integrated moving average~(ARIMA) model.
Median seeing forecast monotonically approaches unconditional median seeing as time $t \rightarrow \infty$ in the model.
The model can be used for time advance up to two hours~\cite{Kornilov2015}.

To obtain $F_{\epsilon}(\epsilon)$ and $Q_{\tau}(p)$ we consider well-known equation from Howell~\cite{Howell2000}:
\begin{equation}
\label{howell_ccd_relative_error}
\frac{1}{\epsilon} = \frac{n\tau}{\sqrt{n\tau + N_{pix}\left(n_s \tau + d \tau + r^2\right)}},
\end{equation}
where $\epsilon$ is the relative photometric error,
$n$ is a total number of photo events for image,
$\tau$ is the exposure time,
$N_{pix}$ is an image size in pixels,
$n_s$ is a photo events from the night sky,
$d$ is a termogeneration ratio,
$r^2$ is a readout noise.
Expression~(\ref{howell_ccd_relative_error}) may be presented in the linear form with respect to this quantities, for instance:
\begin{equation}
\label{ccd_relative_error}
\epsilon^2 = \frac{1}{n \tau} + \frac{\beta^2 s}{n^2 \tau} + \frac{\left(\sigma\beta\right)^2 \left(d \cdot \tau+r^2\right)}{n^2},
\end{equation}
where $\sigma$ is the number of pixels per arc second.
The night sky brightness and the seeing are assumed to obey shifted log-normal distribution~\cite{Kornilov2015}.
From Fenton~\cite{Fenton1960} it follows that the random distributions for $\epsilon^2$ and $\tau$ may be approximated by shifted log-normal distribution.
Then $F_{\epsilon}(\epsilon)$ and $Q_{\tau}(p)$ are corresponding cumulative distribution function and quantile function for log-normal distribution,
the parameters for the distribution are calculated from~(\ref{ccd_relative_error}) using Fenton technique~\cite{Fenton1960}.

\subsubsection{Form of $s(t)$}
Form of $s(t)$ already has been considered in~(\ref{group_task_setup1}) and~(\ref{group_task_setup2}), however the expressions are just recurrent relations.

The setup time $s_{i,j}(t)$ between task $i$ and $j$ is determined by our hardware model with respect to the following reasons:
\begin{itemize}
\item{A telescope mount moving time.
For instance, it takes about $\SI{17}{\second\per\radian}$
for the considering new $\SI{2.5}{\meter}$~telescope of Sternberg Astronomical Institude~(SAI) of Lomonosov Moscow State University~(MSU)~\cite{Kornilov2014} mount to move from one point to another one.
Since that, it follows that common moving time is comparable with common CCD exposure times.
Assuming an acceleration and a deceleration may be neglected the setup time for the mount is the following:
\begin{equation}
\label{mount_setup_time}
s^{\mathrm{Mount}}_{i,j}(t) = C^{\mathrm{Mount}} \max\left\{\left|A_i(t) - A_j(t)\right|,\left|z_i(t) - z_j(t)\right| \right\},
\end{equation}
where $C^{\mathrm{Mount}}$ is an inverse for the velocity, $A_{i,j}(t)$ are azimuths, $z_{i,j}(t)$ are altitudes for $i$ and $j$ targets respectively.
}
\item{Focal port switching.
Some telescopes allows an operator to switch between different focal positions~(Cassegrain, Nasmyth, etc.).
For instance, it takes $\SI{2}{\minute}$ for the third mirror of the $\SI{2.5}{\meter}$ telescope to swap the position selecting one of the five available positions (one Cassegrain position and four Nasmyth ones).
Until now we considered only CCD-based photometry but it is obvious that the described technique may be expanded to different kinds of equipment simultaneously occupying different focal positions.
In general if the task $i$ is performed on the equipment at focal position $o_i$ and the corresponding focal position for the task $j$ is $o_j$; then
\begin{equation}
\label{port_setup_time}
s^{\mathrm{Port}}_{i,j}(t) = C^{\mathrm{Port}} \left(1-\delta_{o_io_j}\right),
\end{equation}
where $C^{\mathrm{Port}}$ is the third mirror swap time, $\delta_{ij}$ denotes generalised Kronecker delta.
}
\item{An internal device state setup.
It is likely that all astronomical equipment have an internal state.
For instance, the state of CCD-based photometer is described by a variety of parameters including the spectral pass band.
In our case the spectral filters are switched by means of rotating a filter wheel, the operation requires about $\SI{10}{\second}$.
So, let $b_i$ is required spectral pass band for the task $i$, then
\begin{equation}
\label{filter_setup_time}
s^{\mathrm{Passband}}_{i,j}(t) = C^{\mathrm{Passband}} \left(1-\delta_{b_ib_j}\right),
\end{equation}
where $C^{\mathrm{Passband}}$ is the constant time to switch the filter position.

This quantity is comparable with minimal available exposure times and well below readout and transmission time which is about one minute.
Unfortunately, E2V~44-82 CCD chip used at the observatory doesn't allow simultaneous readout and exposure, so two consequent CCD-based photometry tasks require the following setup time interval:
\begin{equation}
\label{readout_setup_time}
s^{\mathrm{Readout}}_{i,j}(t) = C_i^{\mathrm{Readout}} N_{pix,i},
\end{equation}
where $C_i^{\mathrm{Readout}}$ is a readout rate and $N_{pix,i}$ is a size of range of interest.
Those parameters are specified by the task.
}
\end{itemize}

All mentioned operations are run in parallel, so the full setup time for the telescope is the following:
\begin{equation}
\label{setup_time}
s_{i,j}(t) = \max\left\{s^{\mathrm{Mount}}_{i,j}(t), s^{\mathrm{Port}}_{i,j}(t),\delta_{o_io_j} \max\left\{s^{\mathrm{Passband}}_{i,j}(t),s^{\mathrm{Readout}}_{i,j}(t)\right\}\right\},
\end{equation}
where $i,j \in T$.
Let us note that expression~(\ref{setup_time}) is not even a metrics due to the term~$s_{ij}^{\mathrm{Readout}}$, which don't obey condition~$s_{ii} \equiv 0$.

Indeed, the model doesn't take into account an overhead affecting both $s_{ij}$ and $d_i$.
The overhead can arise due to controllable reasons like hardware issues which can be solved.
Compare with the atmospheric optical turbulence or sky brightness which can't be excluded in consideration.

\section{Algorithm}
\subsection{Introduction}
\label{algorithm_intro}
Trivially, mean total yield maximisation problem~(\ref{max_mean_yield}) and complete success probability~(\ref{max_prob}) may be formulated as nonlinear integer programming problems using $N^2$ decision variables, where $N \equiv |T|$ is the number of available tasks.
From the papers~\cite{Kannan1978,Belotti2013} it follows that the problems belong to NP-hard class and can't be solved in polynomial time whenever $P \neq NP$~\cite{Leeuwen1990}.

To solve the problems, classical branch-and-bound technique~\cite{Land1960} built on top of parallel depth-bounded discrepancy search~(PDDS) algorithm~\cite{Moisan2014} is used.

The ability to probe on the adequacy of the problem statements is important for us.
We would need a way to check whether the algorithm is appropriate for the problems and whether the problem statements render reality correctly.
We believe that the used approach is more transparent than a neural networks frequently used for similar problems in astronomy~(see the pioneer paper~\cite{Johnston1992}) or genetic algorithms~\cite{Mahoney2012}.
Not to mention that the latter is a subject of fair criticism~(for instance, see the work~\cite{Skiena2008}) and considered as last resort technique.
It is fairly easy to modify used algorithm when initial problem statements~(\ref{max_mean_yield}) or~(\ref{max_prob}) are modified.
For instance, one might introduce dependency constraints in normal disjunctive form.
On the other hand, there is also a freedom to optimise and tune the algorithm.
For instance, more sophisticated heuristics based on a neural networks can be developed.

\begin{figure}[!t]
\centering
\begin{tikzpicture}[
	->,
	xscale=1.0,
	yscale=0.75,
]
\node[align=left] (leafs) at (-6,-3.25) {Leafs};
\node[align=left] (leafs) at (-6,0) {Initial problem};
\node (s1) at (-4,-3.25) {$S^{(1)}$};
\node (s2) at (-3,-3.25) {$S^{(2)}$};
\node (s3) at (-2,-3.25) {$S^{(3)}$};
\node (s4) at (-1,-3.25) {$S^{(4)}$};
\node (s5) at (0,-3.25) {$S^{(5)}$};
\node (s6) at (1,-3.25) {$S^{(6)}$};
\node (s7) at (2,-3.25) {$S^{(7)}$};
\node (s8) at (3,-3.25) {$...$};
\node (s9) at (4,-3.25) {$S^{(|T^+|)}$};
\node (a1) at (-4,-2.5) {} edge (s1);
\node (a2) at (-3,-2.5) {} edge (s2);
\node (a3) at (-2,-2.5) {} edge (s3);
\node (a4) at (-1,-2.5) {} edge (s4);
\node (a5) at (0,-2.5) {} edge (s5);
\node (a6) at (1,-2.5) {} edge (s6);
\node (a7) at (2,-2.5) {} edge (s7);
\node (a8) at (3,-2.5) {} edge (s8);
\node (a9) at (4,-2.5) {} edge (s9);
\node (elipsis) at (0, -2.5) {$...$};
\node (v11) at (-4,-2) {$\left\{i_1i_2\right\}$};
\node (v12) at (-3,-2) {$\left\{i_1i_3\right\}$};
\node (v13) at (-2,-2) {$...$};
\node (v21) at (-1,-2) {$\left\{i_2i_1\right\}$};
\node (v22) at (-0,-2) {$\left\{i_2i_3\right\}$};
\node (v23) at (1,-2) {$...$};
\node (v1) at (-3,-1) {$\left\{i_1\right\}$}
	edge (v11)
	edge (v12)
	edge (v13)
;
\node (v2) at (0,-1) {$\left\{i_2\right\}$}
	edge (v21)
	edge (v22)
	edge (v23)
;
\node (v3) at (3,-1) {$...$};
\node (root) at (0,0) {$\left\{\right\}$}
	edge (v1)
	edge (v2)
	edge (v3)
;
\end{tikzpicture}
\caption{\label{search_tree}
Search tree structure example. Each tree node $V$ is essentially a partial sequence $\{...\}$.
Each edge represents possible appending an item to the partial sequence.
Any tree leaf is a sequence $S \in T^{+}$.
Nothing can be appended further to a leaf.
There are exponentially many leafs in a tree.
Branch and bound technique is used during tree exploration to shrink the search space by ignoring specific sub-trees.
Least discrepancy search technique is used for sorting edges to reach most promising leafs first.
}
\end{figure}
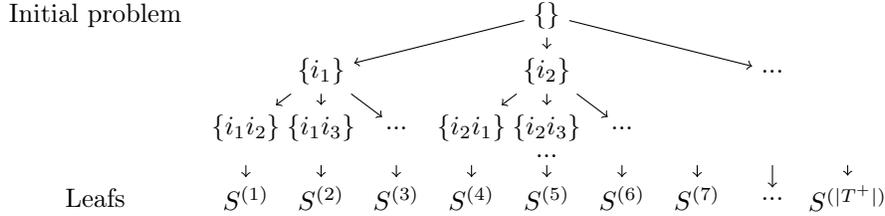

To describe details of the algorithm let us consider a search tree.
An example of such a tree is given in Fig.~\ref{search_tree}.
Any tree leaf corresponds to an element $S \in T^{+}$ of a search space given the constraints.
Any tree node (or vertex) $V$ of depth $D(V)$ corresponds to partial optimisation problem, which is obtained by fixing first $D(V)$ elements. They are $i_1, ..., i_{D(V)}$.
Finally, the tree root corresponds to the initial problem~(\ref{max_mean_yield}) or~(\ref{max_prob}).

PDDS is a parallel modification of depth-first search an essence.
It is a lock-free algorithm where each worker explores dedicated subset.
The subset size difference is $O(N)$.
The authors claim that the algorithm scales well up to thousands of workers due to its lock-free nature~\cite{Moisan2014}.

PDDS is a kind of least discrepancy search~(LDS) algorithm.
An LDS algorithm explores edges in the specific order such that the leafs with the higher probability to reach the maximum are explored first.
It makes possible to reduce the search space dramatically.
Moreover, in practice, if the part of search space is left unexplored; then the algorithm execution time may be arbitrary decreased at the cost of the probability to find global maximum.
For instance, PDDS algorithm has a parameter $k_{max}$, for the nodes with depth greater than $k_{max}$ only single one edge is always explored.
The search order is specified by a local heuristic.
It is a mapping $H$ which introduces an order for the set of all edges $E(V)$ of the vertex $V$.
In practice it may be implemented as an sorting algorithm.

As soon as we consider branch and bound technique, for any tree node we have to define upper bound estimator $B(V)$ which is upper bound for the corresponding partial optimisation problem.
Let $S^{\star}$ denotes the current solution candidate then if $B(V) \le C(S^{\star})$ holds for a node $V$ (where cost function $C$ is either $Y$ or $\Pi$); then $V$-based sub tree doesn't contain a better candidate and must be left unexplored.
Consider now the function $B$.

Note that the following may be used for complete success probability problem~(\ref{max_prob}):
\begin{equation}
\label{bound_prob_unit}
B(V) = \Pi(\left\{i_1,...,i_{D(V)}\right\}).
\end{equation}
Since all $p_i \le 1$ we see that expression~(\ref{prob}) monotonically nonicreases as depth $D(V)$ increases.
Note that problem~(\ref{max_prob}) assumes finding maximum of the upper-bounded function and problem~(\ref{max_prob}) is simpler than mean total yield maximisation problem~(\ref{max_mean_yield}).

However, since function~(\ref{bound_prob_unit}) is too weak on practice let us consider now linear relaxation of Knapsack problem.
Let us recall that Knapsack problem is about packing different discrete items into a knapsack with limited weight capacity.
Each item has two properties: weight and value.
For any item we have to decide whether or not to put it into the knapsack.
The goal is to make overall value of items in the knapsack be highest possible while their weight must not exceed the capacity~\cite{Kellerer2004}.

In our case, the task duration $d_i$ essentially plays the same role as the weight.
The mean yield $y_{i} p_{i}$ and the probability logarithm $\ln p_i$ are counterparts to the item value for~(\ref{max_mean_yield}) and~(\ref{max_prob}).
Indeed, let us rewrite~(\ref{max_prob}) in the following convenient linear form:
\begin{equation}
\label{max_log_prob}
\ln \Pi^* = \max_{S \in T^{+}} \left( \sum_{i=1}^{|S|} \ln\left( p_{S_i}(t_{S_i}) \right) \right).
\end{equation}
Define the notations: $\overline{\ln p_i} \equiv \max_{t} \ln p_i(t)$, $\overline{y_i p_i} \equiv \max_{t} y_i(t) p_i(t)$, $\underline{d_i} \equiv \min_{t} d_i(t)$, $\underline{s_{i,j}} \equiv 0$.
Substituting $\underline{d_i}$ for $d_i$ and $\underline{s_{i,j}}$ and $s_{i,j}$ in~(\ref{max_mean_yield_cst}) and~(\ref{max_prob_cst}) the constraints are kept satisfied for any $S$ and we have the following for any $S^*$:
\begin{equation}
\label{bound_mean_yield_knapsack1}
Y^*(S^*) \le \sum_{i=1}^{|S^*|} \overline{y_{S^*_i} p_{S^*_i}} \le \max_{S \in T^+} \left(\sum_{i=1}^{|S|} \overline{y_{S_i} p_{S_i}} \right),
\end{equation}
\begin{equation}
\label{bound_prob_knapsack1}
\ln \Pi^*(S^*) \le \sum_{i=1}^{|S^*|} \overline{\ln p_{S^*_i}} \le \max_{S \in T^+} \left(\sum_{i=1}^{|S|} \overline{\ln p_{S_i}} \right).
\end{equation}
The right hand side of~(\ref{bound_mean_yield_knapsack1}) with constraints~(\ref{max_mean_yield_cst}) is well-known classic Knapsack problem~\cite{Kellerer2004}.
Similarly, the right hand side of~(\ref{bound_prob_knapsack1}) with constraints~(\ref{max_prob_cst}) is minimal Knapsack problem~\cite{Kellerer2004}.
Note that bounds~(\ref{bound_mean_yield_knapsack1}) and~(\ref{bound_prob_knapsack1}) don't depend on the order inside sequence $S^*$.

There is known $O(N)$\footnote{Naive implementations are $O(N \log N)$.} algorithm to calculate upper bound for Knapsack problem~\cite{Kellerer2004}.
Moreover, it is guaranteed that the upper bound is at most two times greater than optimal value~\cite{Kellerer2004}.
The algorithm is the following.
Ratios ${\overline{y_{S_i} p_{S_i}}} / {\underline{d_{S_i}}}$ are sorted in descending order, then the items are one-by-one included in solution while constraints~(\ref{max_mean_yield_cst}) are satisfied.
The left free space is filled with part of the split item $s$\footnote{The overall complexity is $O(N)$ since that the split item may be found in~$O(N)$~\cite{Kellerer2004}.}, contributing value (yield) proportional to the allocated weight (time).
It is called linear relaxation when a discrete item is allowed to be split.
The similar procedure is used for expression~(\ref{bound_prob_knapsack1}).
Let us denote the upper bound for dominated Knapsack problem for mean total yield maximisation problem~(\ref{max_mean_yield}) and complete success probability maximisation problem~(\ref{max_prob}) as $B_{LR}(V)$.
It is used as the upper bound for the optimisation problems.

The situation is more complicated for the heuristic.
For complete success probability problem~(\ref{max_prob}) sorting the edges descending by the following ratio $e_i$ called an efficiency~\cite{Kellerer2004} appeared to be a good choice.
\begin{equation}
\label{heuristic_prob_knapsack}
e_i = \frac{\ln p_{i}(t_{S_k})}{d_{i}(t_{S_k})},
\end{equation}
I.e. the edge with the maximal ratio is proceeded first.
The heuristic becomes more precise as search depth $k$ increases.

Similar quantity for mean total yield maximisation problem~(\ref{max_mean_yield})
\begin{equation}
\label{heuristic_yield_knapsack}
e_i = \frac{y_{i}(t_{S_k}) p_{i}(t_{S_k})}{d_{i}(t_{S_k})}
\end{equation}
doesn't lead to success,
as well as ascending ordering by the ratio $e'_i$ depending on the derivative over time:
\begin{equation}
\label{heuristic_yield_diff}
e'_i = \left.\frac{\frac{d\left( y_{i}(t) p_{i}(t)\right)}{dt}}{d_{i}(t_{S_k})}\right|_{t = t_{S_k} + \frac{1}{2N}\sum_{j} d_j(t_{S_k})}.
\end{equation}
Expression~(\ref{heuristic_yield_diff}) has been inspired by the paper~\cite{Alidaee1990}.
When seeing conditions are better than average, seeing-demanding CCD-task success probability~(\ref{ccd_task_prob}) decreases,
therefore such tasks having negative $e'_i$ are expected to be placed in the beginning of the sequence.

The function $B_{LR}(V)$ as an edge ordering function appeared to be the most satisfying choice.
The edge with higher upper-bound is proceeded first.

\subsection{Algorithm performance}
The described algorithm and the models have been implemented as a library using C++ programming language\footnote{The reader is referred to \url{https://bitbucket.org/matwey/chelyabinsk} to read the source code.} this allowed us to carry out some performance tests.
Monte-Carlo numeric simulations have been undertaken.
Input data characteristics are described in Table~\ref{solver_run_input}.

\begin{table*}[!t]
\centering
\caption{\label{solver_run_input}
Problem input used for Monte-Carlo simulation.
$U(a,b)$ denotes random uniform distribution.
$LogU(a,b)$ denotes random log-uniform distribution.
}
\begin{tabular}{lrr} 
Parameter & Value & Appearance \\
 & & probability \\
\hline
Right ascending $\alpha$ & $U(\SI{0}{\degree},\SI{360}{\degree})$ & \\
Declination $\delta$ & $U(\SI{-40}{\degree},\SI{90}{\degree})$ & \\
Target flux estimation $n$ & $LogU(-6,7)$ & \\
Relative photometric error $\epsilon$ & $U(0.001,0.1)$ & \\
Yield $y$ & $1$ & \\
Exposure time $\tau$ & $U(\SI{15}{\second},\SI{1500}{\second})$ & $p=0.8$ \\
PSF central intensity $\gamma_1$ & $U(\SI{0.4}{\per\arcsecond{\squared}},\SI{1.2}{\per\arcsecond{\squared}})$ & $p=0.3$ \\
FWHM $\gamma_2$ & $U(\SI{0.4}{\arcsecond},\SI{1.2}{\arcsecond})$ & $p=0.3$ \\
Radius $\gamma_3(e)$ encircling $e$ part of energy & $U(\SI{0.4}{\arcsecond},\SI{1.2}{\arcsecond})$ & $p=0.3$ \\
Part of energy $e$ & $U(0.8,1.0)$ &  \\
Total available observation task number $|T|$ & $30$ & \\
\end{tabular}
\end{table*}

For each problem, $300$ different inputs have been generated and a solution has been found.
There were $30$ available CCD-photometry tasks in each input.
Our estimate of $|T|$ for real application is $70$ per night.
If aggregations from sections~\ref{sss_group_task} and~\ref{sss_repeat_task} are taken into account;
then average task duration is about $\SI{20}{\minute}$ and total required observation time is about $\SI{20}{\hour}$ which is substantial greater than available observational time.

We took $D=\SI{90}{\minute}$ for constraints~(\ref{max_prob_cst}) and~(\ref{max_mean_yield_cst}), which is comparable with available environment forecast limits.
The seeing forecast limit is from one to two hours depending on chosen criterion~\cite{Kornilov2015}.
PDDS $k_{max}$ parameter was chosen to be $8$.
In average, $|S|$ appeared to be $10$.
We can estimate total number of possible combinations as $|T^{+}| > \num{1e15}$, which would have required few CPU-months for brute-force algorithm.

\begin{figure}[!t]
\centering
\includegraphics[width=75mm]{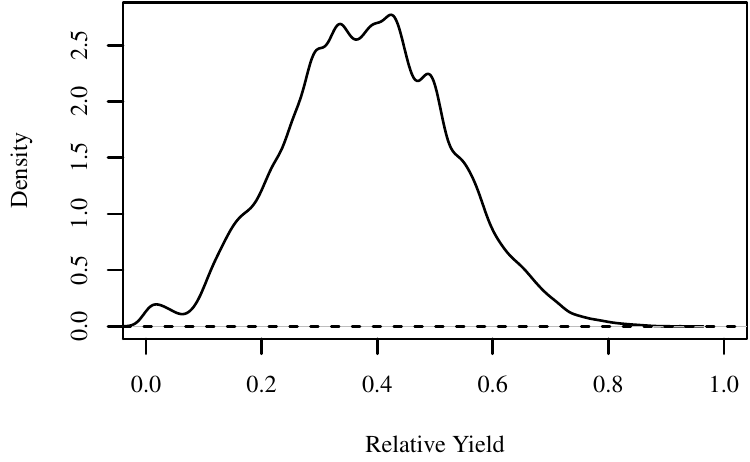}
\caption{\label{yield_random} Probability density function for ${Y(S)}/{Y^{*}}$ when $S$ are uniformly distributed.}
\end{figure}

Sample probability density function for ${Y(S)}/{Y^{*}} \le 1$ is given in Fig.~\ref{yield_random}, here samples $S$ obey uniform random distribution.
Let us recall, that $Y^{*}$ denotes the optimal value of the target function.
If the distribution is assumed to be normal one; then the probability that the solution is found by chance is less than $\num{1e-5}$.
Fortunately, proposed algorithm is way faster and we carried out simulation and discuss the results here.

\begin{figure}[!t]
\centering
\includegraphics[width=75mm]{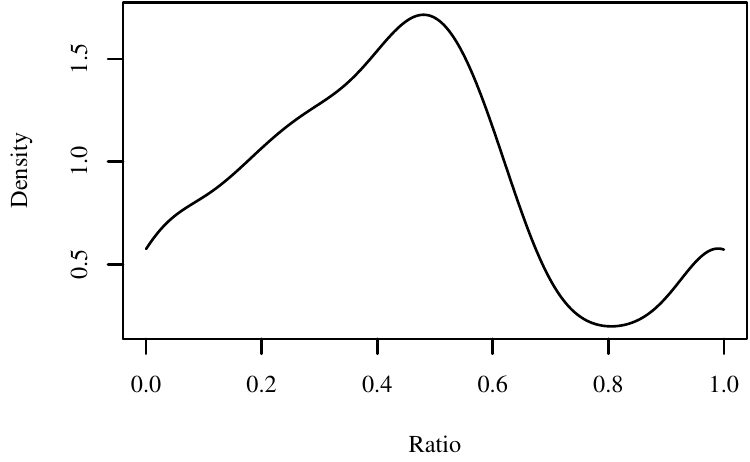}
\caption{\label{prob_bound_ratio} Ratio between $B_{LR}(V_0)$ and found solution for success probability maximisation problem~(\ref{max_log_prob}).
The problem target function is negative, so $|B_{LR}(V_0)| < |\ln \Pi^{*}|$ and the ratio is less than $1$.}
\end{figure}

\begin{figure}[!t]
\centering
\includegraphics[width=75mm]{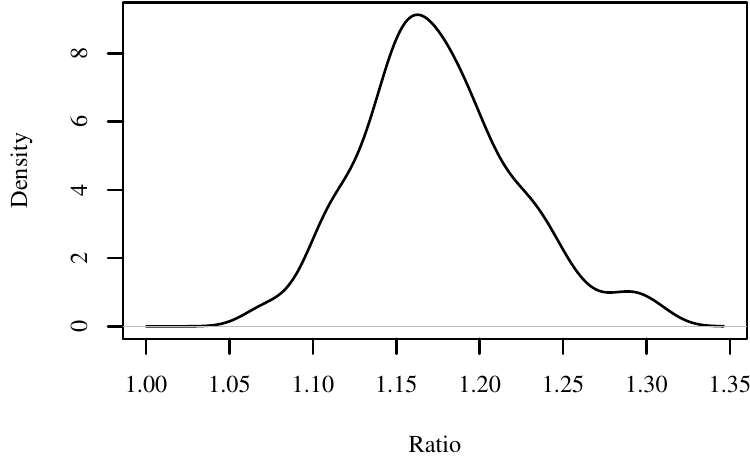}
\caption{\label{yield_bound_ratio} Ratio between $B_{LR}(V_0)$ and found solution for mean total yield maximisation problem~(\ref{max_mean_yield}).}
\end{figure}

An efficiency of upper-bound functions $B_{LR}(V_0)$ for the first level is demonstrated in Fig.~\ref{prob_bound_ratio} and~\ref{yield_bound_ratio}. 
The closer value to $1$, the smaller search space.
Since the solution are bounded twice in~(\ref{bound_mean_yield_knapsack1}) and~(\ref{bound_prob_knapsack1}), we can't give any theoretical estimator for this ratio.

Fig.~\ref{yield_bound_ratio} may be also interpreted as the following.
The found solution for mean total yield maximisation problem~(\ref{max_mean_yield}) is $1.15$ times less than trivially calculated upper bound in average.
This statement allows us to estimate the precision scale of the optimisation problem solution.

\begin{figure}[!t]
\centering
\includegraphics[width=75mm]{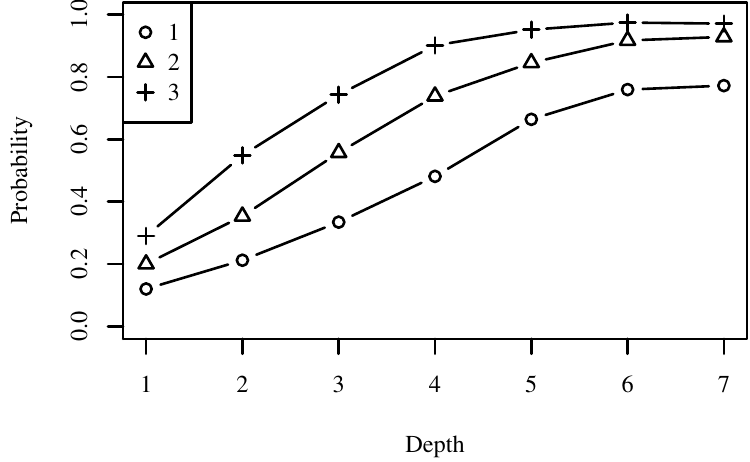}
\caption{\label{prob_h_ks} Probability that the problem solution is located within first $N$ ($N = 1,2,3$) edges as a function of level $k$.
Success probability maximisation problem~(\ref{max_log_prob}) and heuristic~(\ref{heuristic_prob_knapsack}).}
\end{figure}

\begin{figure}[!t]
\centering
\includegraphics[width=75mm]{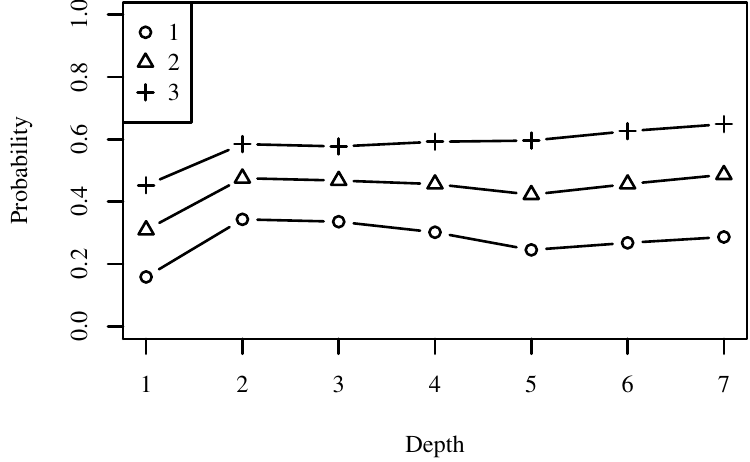}
\caption{\label{yield_h_ks} Probability that the problem solution is located within first $N$ ($N = 1,2,3$) edges as a function of level $k$.
Mean total yield maximisation problem~(\ref{max_mean_yield}) and heuristic~(\ref{heuristic_yield_knapsack}).}
\end{figure}

\begin{figure}[!t]
\centering
\includegraphics[width=75mm]{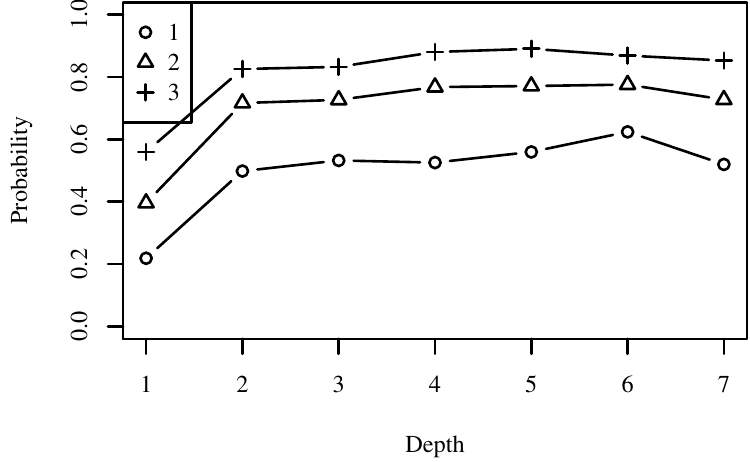} 
\caption{\label{yield_h_bound} Probability that the problem solution is located within first $N$ ($N = 1,2,3$) edges as a function of level $k$.
Mean total yield maximisation problem~(\ref{max_mean_yield}) and heuristic~$B_{LR}(V)$.}
\end{figure}

The behaviour of the heuristic for complete success probability problem~(\ref{max_log_prob}) is given in Fig.~\ref{prob_h_ks}.
Evidently, that the heuristic misses less often as depth increases.
This is the base for PDDS algorithm~\cite{Moisan2014}.
The similar heuristic for problem~(\ref{max_mean_yield}) performs worse, as one can see in Fig.~\ref{yield_h_ks} and~\ref{yield_h_bound}.

Let us recall that we assume that $p(t)$ in~(\ref{max_log_prob}) and~(\ref{max_mean_yield}) are conditional probabilities $p(t|\theta_0)$ in fact.
The current system state is denoted by $\theta_0$.
For the future time moment $t'_0 > t_0$ the following expression can be written:
\begin{equation}
p(t|\theta_0)=\int d\omega p(t|\theta'_0; \omega) p(\theta'_0|\theta_0; \omega),
\end{equation}
where $\omega$ is a member of sample space for system evolution from $t_0$ to $t'_0$.
Thereby, $p(t|\theta_0)$ may be considered as the mean for $p(t|\theta)$ averaged over all possible system evolutions.

The following numerical simulation was carried out in order to estimate the solution stability with respect to the system evolution for mean total yield maximisation problem~(\ref{max_mean_yield}).
For each of total 300 tests, $S^*$ has been found and the optical turbulence evolution has been modelled until the time $t_2$.
To simulate turbulence evolution, we use the same ARIMA-based model which is mentioned in section~\ref{ccd_based_photometry}~\cite{Kornilov2015}.
Let $S^0$ denote $S^*$ without the first element, and let $S^1$ denote solution for the problem with the modified initial state.
It simulates the scheduling rerun at the time moment $t_2$.
In the simulations, $t_2$ appeared to be $\SI{215}{\second}$ in average.

\begin{figure}[!t]
\centering
\includegraphics[width=75mm]{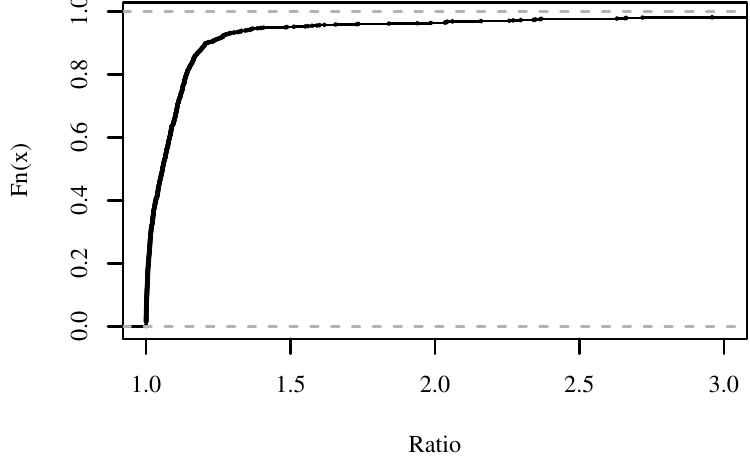}
\caption{\label{otvar_ratio} Cumulative distribution function for ratio ${Y(S^0)}/{Y(S^1)}$.}
\end{figure}

\begin{figure}[!t]
\centering
\includegraphics[width=75mm]{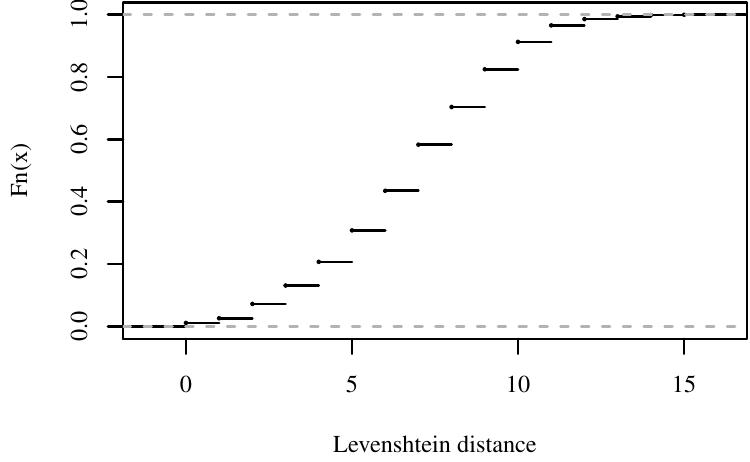}
\caption{\label{otvar_dist} Cumulative distribution function for Levenshtein metrics between $S^0$ and $S^1$.}
\end{figure}

\begin{figure}[!t]
\centering
\includegraphics[width=75mm]{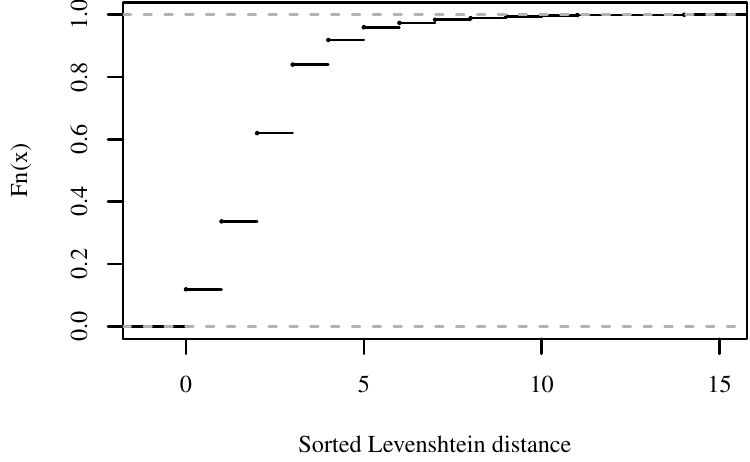} 
\caption{\label{otvar_dist2} Cumulative distribution function for Levenshtein metrics between sorted $\Theta(S^0)$ and $\Theta(S^1)$.}
\end{figure}

\begin{figure}[!t]
\centering
\includegraphics[width=75mm]{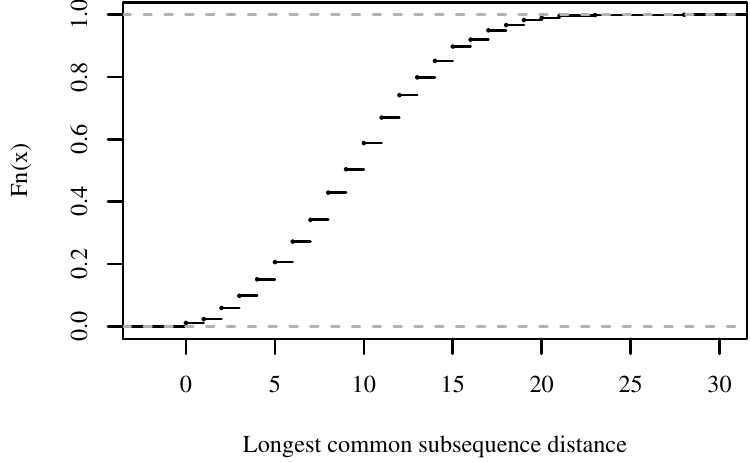}
\caption{\label{otvar_dist3} Cumulative distribution function for longest common subsequence metrics between $S^0$ and $S^1$.}
\end{figure}

The cumulative distribution function for the ratio ${Y(S^1)}/{Y(S^0)}$ is given in Fig.~\ref{otvar_ratio}.
The cumulative distribution function for the different distance metrics between $S^1$ and $S^0$ are shown in Fig.~\ref{otvar_dist} and~\ref{otvar_dist2}.
The Levenshtein distance was chosen as metrics.~\cite{Levenshtein1966}.
In Fig.~\ref{otvar_dist2}, the distance between $\Theta(S^0)$ and $\Theta(S^1)$ is considered, where $\Theta$ is some sorting rule.
This way, the number of different items in $S^0$ and $S^1$ is considered in Fig.~\ref{otvar_dist2}.
The common longest subsequence metrics is considered in Fig.~\ref{otvar_dist3}.
One may see, that the item order are most unstable, but the content of $S$ and the target function value rather remain the same.

The conclusion is not quite optimistic because the reason the task order is optimised is that we want to optimise the cumulative setup time.
It requires some degree of stability.
If we refuse the idea to optimise the order; then the next observational task can be selected by one of the considered heuristics for each time moment $t$.
The local greedy approach is a way simpler than described discrete optimisation from practical point of view.

If we want to keep the global optimisation approach; then to restate mean total yield maximisation problem~(\ref{max_mean_yield}) and complete success probability maximisation problem~(\ref{max_prob}) taking a variance of $p(t)$ into account could be possible solution.
Let $\sigma^2(t|\theta_0)$ denote the variance of $p(t) \equiv p(t|\theta_0)$ with respect to all possible system evolutions from $t_0$ to $t'_0$. Then we have:
\begin{equation}
\label{p_variance}
\sigma^2(t|\theta_0) + p^2(t|\theta_0) =\int d\omega p^2(t|\theta'_0; \omega) p(\theta'_0|\theta_0; \omega) = p(t|\theta'_0;\omega^*)p(t|\theta_0),
\end{equation}
From~(\ref{p_variance}) it follows that $\sigma^2(t|\theta_0) \rightarrow 0$ as $p(t|\theta_0) \rightarrow 0$.
Moreover, since $p(t|\theta_0) \le p(t|\theta'_0;\omega^*) \le 1$ we see that  $\sigma^2(t|\theta_0) \rightarrow 0$ as $p(t|\theta_0) \rightarrow 1$.
Hence, the probability that the probability $p(t)$ changes considerably is low if the probability $p(t)$ itself is great regardless of the specific form and distribution of $p(t)$.
In general, a moderate $p(t)$ should not be placed far from the sequence $S$ begin.
However, this is a subject of further research.

\begin{figure}[!t]
\centering
\includegraphics[width=75mm]{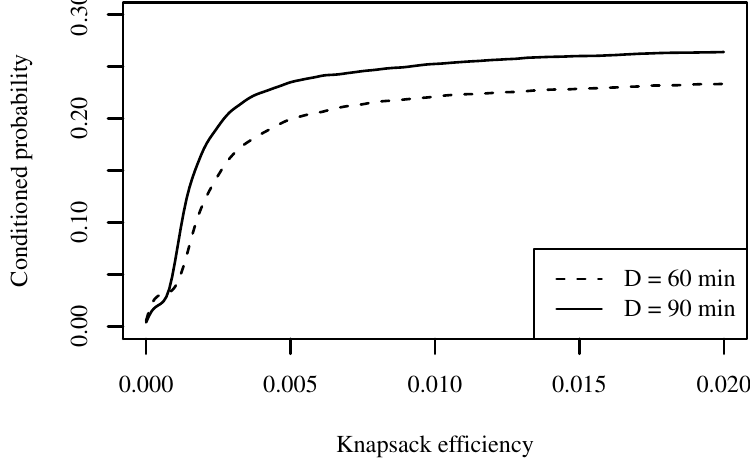}
\caption{\label{pi_cond_e} Conditional probability $F_{\pi,e}(1|e)$ that observational task gets into the solution given efficiency $e$
where $e$ is efficiency from~(\ref{heuristic_yield_knapsack}).}
\end{figure}

A conditional probability $F_{\pi,e}(1|e)$ that the observational task gets into the solution given $e$ is given in Fig.~\ref{pi_cond_e}.
Recall, that $e$ is defined in~(\ref{heuristic_yield_knapsack}).
For instance, if $d=\SI{50}{\second}$, $p=1$, $y=1$; then the efficiency $e=\SI{0.02}{\per\second}$.
The form of the curve in Fig.~\ref{pi_cond_e} is well approximated by $C_1-C_2 \ln(e) e^{-1}$ except the small area near zero.
The corresponding conditional probability density $p_{\pi,e}(1|e)$ may be obtained as the following:
\begin{equation}
\label{cond_prob}
p_{\pi,e}(1|e) = F_{\pi,e}(1|e) + e\frac{\partial F_{\pi,e}(1|e)}{\partial e},
\end{equation}
and generally follows the behaviour of $F_{\pi,e}(1|e)$.

The explanation why heuristic~(\ref{heuristic_yield_knapsack}) is so inefficient follows from Fig.~\ref{pi_cond_e}.
For any task such that its efficiency $e$ is greater than some limit it follows that the probability to appear in the solution just a little depends on $e$.

Of course, the specific function form depends on a problem statement: an input task set, a current state $\theta_0$ determining $p_i(t|\theta_0)$, and the  constraint $D$.
For instance, the asymptotic is determined by the ratio between number of available tasks and the available time resources $D$.
However, now we see that $y(t)$ are defined on some nonlinear generally unknown scale if we consider the frequency of occurrence of the task is in the solution. 
Let two tasks $i$ and $j$ be called $p_e$-equivalent iff $e_i = e_j$, that is they have the same probability to appear in the solution.
This definition doesn't depend on the specific form of $p_{\pi,e}(1|e)$.
Assuming that $p_{\pi,e}(1|e)$ is monotonic, we can introduce an order on the available task set $T$ called $p_e$-order for any particular set of $p_i$ and $d_i$.

A telescope scientific committee may introduce an explicit order on the available task set $T$ making at most $|T|\ln_2|T|$ binary decisions whether the task $i$ is more important that $j$ or not.
The order can be made to coincide with $p_e$-order with average $p_i$ and $d_i$ by assigning the specific yields $y_i$.
Note, that it is only one from variety of possible ways to assign yield $y_i$.
Unfortunately, a concept of scientific importance is widely adopted in works on automatic scheduling (for instance see~\cite{Gomez2003}), in spite of all scientific knowledge is considered to be equally valuable by modern philosophy of science.

Another, more fair way to assign $y_i$ could be based on unconditional success probabilities $\hat p_i = \int p_i(t|\theta)p(t,\theta)d\theta dt$.
Environment parameters are specifically distributed, so different tasks have different amount of appropriate time to be carried out.
If all tasks are $p_{\hat e}$-equivalent with respect to the unconditionally mean $\hat e$ then we call their yields $y$ as unconditionally fair yields.

Yet another interpretation for $y_i$ is related to open shutter time that is just $\sum_{i=1}^{|S|} d_{S_i} (t_{S_i})$ in our terms.
We define net open shutter time as the following:
\begin{equation}
\label{stochastic_ost}
{\mathcal O} \equiv \sum_{i=1}^{|S|} d_{S_i}(t_{S_i}) \xi_{S_i}(t_{S_i}),
\end{equation}
where $\xi_{S_i}(t_{S_i})$ are random binary variables being~$1$ with probability of $p_{S_i}(t_{S_i})$.
Unlike the open shutter time, quantity ${\mathcal O}$ includes only successful tasks.
Its mean is expressed as the following:
\begin{equation}
\label{mean_ost}
O = \sum_{i=1}^{|S|} d_{S_i}(t_{S_i}) p_{S_i}(t_{S_i}).
\end{equation}
One may see that the mean net open shutter time $O$ is proportional to the mean total yield when $y_i(t) \equiv C^{-1} \cdot d_i(t)$ for any $i$,
where constant $C$ may be chosen as $\max_{i,t} d_i(t)$.

\begin{figure*}[!t]
\centering
\includegraphics[width=140mm]{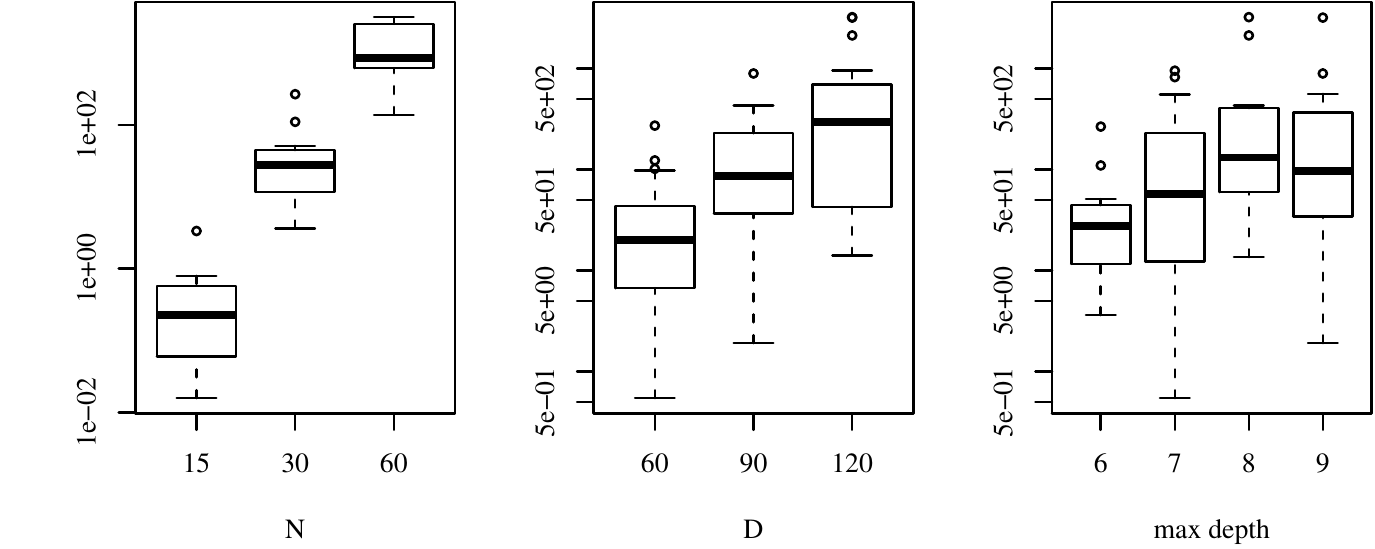}
\caption{\label{runtime} Box-and-whisker plot for solving mean total yield maximisation problem~(\ref{max_mean_yield}) run time in seconds.
Bottoms and tops of boxes are first and third quartiles.
Solid lines are medians.
Individual circles are outliers, which are out of $1.5$ inter quartile range.
Whiskers denote sample minimum and maximum (except outliers).
Different parameters are amount of available tasks $N$, scheduling time horizon $D$ (minutes), PDDS $k_{max}$ limit depth parameter.}
\end{figure*}

The run times with different parameters (the time constraint constant $D$, amount of input tasks $N$, and $k_{max}$ parameter~\cite{Moisan2014}) are demonstrated in Fig.~\ref{runtime}.
The solutions were found using four threads on four cores of commercial CPU Xeon E5-2630L.
It should be understood, that the tests are rather synthetic.
No doubts that there are a lot possibilities to optimise the algorithm.
However, testing running at real telescope will probably highlight another issues arising with the real world input data.

\section{Conclusion}
Aspects of online automatic observation scheduling have been considered.
The mean total yield $Y$ and the complete success probability $\Pi$ have been defined and have been connected with the concept of efficiency.
Mean total yield $Y$ maximisation problem~(\ref{max_mean_yield}) and success probability $\Pi$ maximisation problem~(\ref{max_prob}) have been stated.
These discrete optimisation problems are considered as efficiency maximising ones.

The a-priory success probabilities introduced in section~\ref{opt_prob} require the models of the physical underlying experimental equipment.
The problem functions $p(t)$, $d(t)$ и $s(t)$ form the abstraction layer between the equipment model and the optimisation problem.
The instance of hardware model for CCD-based photometry has been described,
but it is also possible to develop models for other astronomical equipment.

The probabilistic equipment model is based on environment models, for instance, the optical turbulence forecast model~\cite{Kornilov2015} and the night sky brightness model.
An input for the environment models is supposed to be obtained online using dedicated monitoring experiments, for instance, the capable automatic seeing monitor has been running on the site of the SAI MSU $\SI{2.5}{\meter}$ telescope for eight years.~\cite{Kornilov2014}.
This way, the optimisation problems are to be solved a number of times per night using actual current initial state.
On the other hand, the available observation task set $T$ can also be varied between scheduling runs,
for instance in case of transient objects like gamma ray bursts.
However, it is out of present paper scope and requires further development,
because such targets may have considerable uncertainties in their position and other parameters which have to be properly accounted for by success probability function $p(t)$.

To solve mean total yield maximisation problem~(\ref{max_mean_yield}) and success probability maximisation problem~(\ref{max_prob}), the PDDS algorithm~\cite{Moisan2014} is used.
The algorithm is generally capable to be used for online astronomical scheduling.
The approach seems to have a lot of possibilities for further development and tuning.

\bibliographystyle{elsarticle-num}
\bibliography{biblio}

\begin{thebibliography}{10}
\expandafter\ifx\csname url\endcsname\relax
  \def\url#1{\texttt{#1}}\fi
\expandafter\ifx\csname urlprefix\endcsname\relax\def\urlprefix{URL }\fi
\expandafter\ifx\csname href\endcsname\relax
  \def\href#1#2{#2} \def\path#1{#1}\fi

\bibitem{Moisan2014}
T.~{Moisan}, C.-G. {Quimper}, J.~{Gaudreault}, Parallel depth-bounded
  discrepancy search, in: H.~Simonis (Ed.), Integration of AI and OR Techniques
  in Constraint Programming, Vol. 8451 of Lecture Notes in Computer Science,
  Springer International Publishing, 2014, pp. 377--393.
\newblock \href {http://dx.doi.org/10.1007/978-3-319-07046-9\_27}
  {\path{doi:10.1007/978-3-319-07046-9\_27}}.

\bibitem{Bowen1964}
I.~S. {Bowen}, {Telescopes}, Astronomical Journal 69 (1964) 816.
\newblock \href {http://dx.doi.org/10.1086/109358} {\path{doi:10.1086/109358}}.

\bibitem{Gomez2003}
{A. I. {Gómez de Castro}}, {J. {Yáñez}}, Optimization of telescope
  scheduling, Astronomy and Astrophysics 403~(1) (2003) 357--367.
\newblock \href {http://dx.doi.org/10.1051/0004-6361:20030319}
  {\path{doi:10.1051/0004-6361:20030319}}.

\bibitem{Colome2010}
J.~{Colomé}, K.~{Casteels}, I.~{Ribas}, X.~{Francisco}, The {TJO-OAdM} robotic
  observatory: the scheduler, in: Proc. SPIE, Vol. 7740, 2010, pp.
  77403K--77403K--12.
\newblock \href {http://dx.doi.org/10.1117/12.857672}
  {\path{doi:10.1117/12.857672}}.

\bibitem{Delgado2014}
F.~{Delgado}, G.~{Schumacher}, The {LSST} {OCS} scheduler design, in: Proc.
  SPIE, Vol. 9149, 2014, pp. 91490G--91490G--13.
\newblock \href {http://dx.doi.org/10.1117/12.2056871}
  {\path{doi:10.1117/12.2056871}}.

\bibitem{Kornilov2015}
M.~V. {Kornilov}, {Forecasting seeing and parameters of long-exposure images by
  means of ARIMA}, Experimental Astronomy 41 (2016) 223--242.
\newblock \href {http://dx.doi.org/10.1007/s10686-015-9485-7}
  {\path{doi:10.1007/s10686-015-9485-7}}.

\bibitem{Howell2000}
S.~B. {Howell}, {Handbook of CCD Astronomy}, 2nd Edition, Cambridge University
  Press, 2000.

\bibitem{Fenton1960}
L.~Fenton, The sum of log-normal probability distributions in scatter
  transmission systems, IRE Trans. Commun. Systems 8~(1) (1960) 57--67.
\newblock \href {http://dx.doi.org/10.1109/TCOM.1960.1097606}
  {\path{doi:10.1109/TCOM.1960.1097606}}.

\bibitem{Kornilov2014}
V.~{Kornilov}, B.~{Safonov}, M.~{Kornilov}, N.~{Shatsky}, O.~{Voziakova},
  S.~{Potanin}, I.~{Gorbunov}, V.~{Senik}, D.~{Cheryasov}, {Study on
  Atmospheric Optical Turbulence above Mount Shatdzhatmaz in 2007-2013},
  Publications of the ASP 126 (2014) 482--495.
\newblock \href {http://dx.doi.org/10.1086/676648} {\path{doi:10.1086/676648}}.

\bibitem{Kannan1978}
R.~{Kannan}, C.~L. {Monma}, On the computational complexity of integer
  programming problems, in: R.~Henn, B.~Korte, W.~Oettli (Eds.), Optimization
  and Operations Research, Vol. 157 of Lecture Notes in Economics and
  Mathematical Systems, Springer Berlin Heidelberg, 1978, pp. 161--172.
\newblock \href {http://dx.doi.org/10.1007/978-3-642-95322-4\_17}
  {\path{doi:10.1007/978-3-642-95322-4\_17}}.

\bibitem{Belotti2013}
P.~{Belotti}, C.~{Kirches}, S.~{Leyffer}, J.~{Linderoth}, J.~{Luedtke},
  A.~{Mahajan}, Mixed-integer nonlinear optimization, Acta Numerica 22 (2013)
  1--131.
\newblock \href {http://dx.doi.org/10.1017/S0962492913000032}
  {\path{doi:10.1017/S0962492913000032}}.

\bibitem{Leeuwen1990}
J.~{van Leeuwen} (Ed.), Handbook of Theoretical Computer Science, Volume {A:}
  Algorithms and Complexity, Elsevier and {MIT} Press, 1990.

\bibitem{Land1960}
A.~H. {Land}, A.~G. {Doig}, An automatic method of solving discrete programming
  problems, Econometrica 28~(3) (1960) 497--520.

\bibitem{Johnston1992}
M.~{Johnston}, H.-M. {Adorf}, Scheduling with neural networks—the case of the
  hubble space telescope, Computers and Operations Research 19~(3) (1992) 209
  -- 240.
\newblock \href {http://dx.doi.org/10.1016/0305-0548(92)90045-7}
  {\path{doi:10.1016/0305-0548(92)90045-7}}.

\bibitem{Mahoney2012}
W.~{Mahoney}, C.~{Veillet}, K.~{Thanjavur}, A genetic algorithm for
  ground-based telescope observation scheduling, in: Proc. SPIE, Vol. 8448,
  2012, pp. 84480W--84480W--14.
\newblock \href {http://dx.doi.org/10.1117/12.926662}
  {\path{doi:10.1117/12.926662}}.

\bibitem{Skiena2008}
S.~S. {Skiena}, The Algorithm Design Manual, 2nd Edition, Springer Publishing
  Company, Incorporated, 2008.

\bibitem{Kellerer2004}
H.~{Kellerer}, U.~{Pferschy}, D.~{Pisinger}, Knapsack Problems, Springer,
  Berlin, Germany, 2004.

\bibitem{Alidaee1990}
B.~{Alidaee}, A heuristic solution procedure to minimize makespan on a single
  machine with non-linear cost functions, The Journal of the Operational
  Research Society 41~(11) (1990) 1065--1068.

\bibitem{Levenshtein1966}
V.~I. {Levenshtein}, Binary codes capable of correcting deletions, insertions
  and reversals, Soviet Physics Doklady 10 (1966) 707.

\end{thebibliography}

\end{document}